\documentclass[12pt]{article}

\usepackage[utf8]{inputenc}
\usepackage{alphabeta}
\usepackage{amsmath}
\usepackage{amssymb}
\usepackage{mathrsfs}
\usepackage{graphicx}
\usepackage{bbold}
\usepackage{comment}
\usepackage{transparent}
\usepackage{bold-extra}
\usepackage[dvipsnames]{xcolor}
\usepackage[most]{tcolorbox}

\usepackage{braket}
\usepackage{bbm}
\usepackage{physics}
\usepackage{comment}
\usepackage{listings}

\numberwithin{equation}{section}

\renewcommand{\lstlistingname}{\bfseries Listing}
\makeatletter
\def\fnum@lstlisting{%
  \lstlistingname
  \ifx\lst@@caption\@empty\else~\thelstlisting\normalfont\fi}%
\makeatother

\definecolor{codegreen}{rgb}{0,0.6,0}
\definecolor{codegray}{rgb}{0.5,0.5,0.5}
\definecolor{codepurple}{rgb}{0.58,0,0.82}
\definecolor{backcolour}{rgb}{0.95,0.95,0.92}
\definecolor{doccolour}{rgb}{0.92,0.94,0.96}

\lstdefinestyle{mystyle}{
    backgroundcolor=\color{backcolour},   
    commentstyle=\color{codegreen},
    keywordstyle=\color{magenta},
    numberstyle=\tiny\color{codegray},
    stringstyle=\color{codepurple},
    basicstyle=\ttfamily\footnotesize,
    breakatwhitespace=false,         
    breaklines=true,                 
    captionpos=b,                    
    keepspaces=true,                 
    numbers=left,                    
    numbersep=5pt,                  
    showspaces=false,                
    showstringspaces=false,
    showtabs=false,                  
    tabsize=2
}

\renewcommand{\baselinestretch}{1.2}
\newcommand{\del}{\partial}

\usepackage{braket}
\usepackage{bm}
\usepackage{bbm}
\usepackage{hyperref}
\numberwithin{equation}{section}

\newcommand{\mat}[1]{\left(\begin{matrix}#1\end{matrix}\right)}
\newcommand{\bat}[1]{\begin{matrix}#1\end{matrix}}
\newcommand{\rep}[2]{\textbf{#1}_\textbf{#2}}
\newcommand{\ihat}{\hat\imath}
\newcommand{\bs}{\boldsymbol}

\newcommand{\underlabel}[2]{\underset{#1}{\underbrace{#2}}}
\newcommand{\overlabel}[2]{\overset{#1}{\overbrace{#2}}}

\newcommand {\nn} {\nonumber}
\newcommand{\bbR}{{\mathbb R}}
\newcommand{\bbC}{{\mathbb C}}
\newcommand{\mycomment}[1]{}

\usepackage{subfig}
\makeatother

\begin{document}
\begin{titlepage}
\renewcommand{\thefootnote}{\fnsymbol{footnote}}

\begin{flushright} 
  
\end{flushright} 

\vspace{1.5cm}

\begin{center}
  {\bf \large
  Reduced tensor network formulation for non-Abelian gauge theories in arbitrary dimensions
  }
\end{center}

\vspace{1cm}


\begin{center}
         Atis Y{\sc osprakob}\footnote
          { E-mail address : ayosp(at)phys.sc.niigata-u.ac.jp}


\vspace{1cm}

\textit{Department of Physics, Niigata University, Niigata 950-2181, Japan}

\end{center}

\vspace{0.5cm}

\begin{abstract}
  \noindent

Formulating non-Abelian gauge theories as a tensor network is known to be challenging due to the internal degrees of freedom that result in the degeneracy in the singular value spectrum. In two dimensions, it is straightforward to `trace out' these degrees of freedom with the use of character expansion, giving a reduced tensor network where the degeneracy associated with the internal symmetry is eliminated. In this work, we show that such an index loop also exists in higher dimensions in the form of a closed tensor network we call the `armillary sphere'. This allows us to completely eliminate the matrix indices and reduce the overall size of the tensors in the same way as is possible in two dimensions. This formulation allows us to include significantly more representations with the same tensor size, thus making it possible to reach a greater level of numerical accuracy in the tensor renormalization group computations.

\end{abstract}
\vfill
\end{titlepage}
\vfil\eject


\renewcommand{\thefootnote}{\arabic{footnote}}
\setcounter{footnote}{0}

\tableofcontents

\section{Introduction}

The tensor network method is a promising approach in the investigation of gauge theory because it is free from the sign problem by definition and because of its accessibility to large volumes at a logarithmic cost \cite{Levin:2006jai,PhysRevLett.115.180405,Adachi:2020upk,PhysRevB.86.045139,Adachi:2019paf,Kadoh:2019kqk,Sakai:2017jwp,Gu:2013gba,Akiyama:2020sfo}. It is also possible to study the fermionic systems directly without the need to integrate the Grassmann variables beforehand \cite{Gu:2010yh,Gu:2013gba,Shimizu:2014uva,Akiyama:2020sfo,Sakai:2017jwp,Yosprakob:2023flr}. This allows us to study many interesting systems and parameter regions, including those that are difficult for the Monte Carlo methods. Notable examples include two-dimensional gauge theories with a $\theta$ term \cite{Kuramashi:2019cgs,Fukuma:2021cni,Hirasawa:2021qvh}, 2D SU(2) gauge-Higgs model \cite{Bazavov:2019qih}, one-flavor Schwinger model \cite{Shimizu:2014uva,Shimizu:2014fsa,Shimizu:2017onf}, $N_f$-flavor gauge theories \cite{Yosprakob:2023tyr}, 2D QCD \cite{Bloch:2022vqz}, 3D SU(2) gauge theory \cite{Kuwahara:2022ald}, 4D $\mathbb{Z}_N$ gauge-Higgs models \cite{Akiyama:2022eip,Akiyama:2023hvt} and many others.

However, the current formulation of non-Abelian gauge theory as a tensor network still suffers from various problems. On one hand, it is possible to easily construct the initial tensor using the sampling method, where the group integration of the link variables is approximated by a Monte Carlo integration \cite{Fukuma:2021cni,Kuwahara:2022ald}. However, the performance of this method directly depends on the choice of the trial action, which can be highly non-trivial to choose. It is also known that the singular value spectrum for non-Abelian gauge theory suffers from severe degeneracy, which has been shown explicitly in two dimensions \cite{Fukuma:2021cni}. The cause of this can be seen when using the character expansion---the degeneracy can be attributed to the internal symmetry of different representations of the gauge group.

In the tensor network construction of a non-Abelian gauge theory in two dimensions using character expansion, the matrix indices for each representation form a closed loop around each lattice site and can be traced out analytically \cite{Hirasawa:2021qvh}. The tensor leg, which originally consisted of a representation index and two matrix indices, can thus be reduced to just a single representation index. In such a reduced tensor network, the degeneracy associated with the internal symmetry no longer appears in the spectrum.
Although the tensor network construction based on character expansion in 3 dimensions has been previously proposed \cite{Liu:2013nsa}, the possibility of tracing out the matrix indices has not been discussed.

In this paper, we show that a similar reduction can be performed in any dimension. The core of the idea is that it is possible to decompose the integral of the link variables into two vertex tensors.
These vertex tensors can be shown to always form a closed structure around each lattice site, which we refer to as the `armillary sphere,' where the matrix indices are completely contracted. This is the generalization of the index loops in two dimensions. This is equivalent to the spin-foam formulation of the pure non-Abelian gauge theory introduced in Ref. \cite{Oeckl:2000hs}. After the contraction of matrix indices, the tensor network now consists of tensors whose legs are only associated with the representation indices and no matrix indices. We also discuss the extension of our idea to a more general Wilson loop action.

This paper is organized as follows. In section \ref{sec:formulation}, we give the construction of the tensor network for the non-Abelian gauge theory in $d$ dimensions using character expansion. In section \ref{sec:loops}, we discuss how to evaluate the group integral via the generalized Clebsch-Gordan coefficients. The link integral is then shown to be decomposable into two vertex tensors. We also define the armillary sphere as a closed structure of the vertex tensors. The idea is then generalized to other Wilson loop actions, including the theta term. Section \ref{sec:summary} is devoted to the summary and discussions.

\section{Tensor network formulation of gauge theory}
\label{sec:formulation}
\begin{figure}
    \centering
    \includegraphics[scale=0.8]{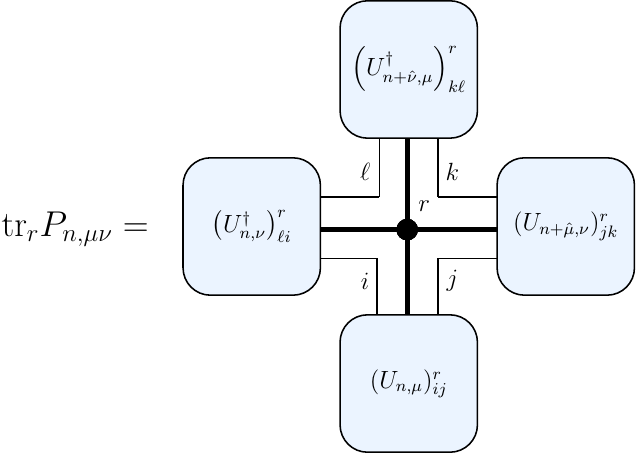}
    \caption{The tensor diagram of a plaquette trace in the $r$-representation (the $r$-character). Thin lines are matrix indices while the thick lines are representation indices. The node in the middle is the Kronecker delta tensor forcing the 4 link variables to be under the same representation.}
    \label{fig:plaquette}
\end{figure}

Consider a gauge theory with the gauge group $G$ on a $d$-dimensional hyper-cubic lattice $\Lambda_d$. The site is denoted by $n$ and the link variable between the site $n$ and $ n+\hat\mu$ is given by a matrix $U_{n,\mu}\in G$, which we assume to be under a unitary representation of $G$. A plaquette variable $P_{n,\mu\nu}$
is defined by
\begin{equation}
    P_{n,\mu\nu}\equiv U_{n,\mu}U_{n+\hat\mu,\nu}U^\dagger_{n+\hat\nu,\mu}U^\dagger_{n,\nu}.
    \label{eq:plaquette}
\end{equation}
We will now consider the partition function of a pure gauge theory
\begin{align}
    Z &= \int DU e^{-S[U]};\\ 
    S[U] &= -\beta\sum_{n\in\Lambda_d}\sum_{1\leq\mu<\nu\leq d} \Re\tr P_{n,\mu\nu}.
\end{align}
Using character expansion \cite{Hirasawa:2021qvh,Bars:1979xb,Bars:1980yy,Samuel:1980vk}, the Boltzmann weight associated with a plaquette $P_{n,\mu\nu}$ is given by
\begin{align}
    e^{\beta \Re\tr P_{n,\mu\nu}} &= \sum_r f_r \tr_r(U_{n,\mu}U_{n+\hat\mu,\nu}U^\dagger_{n+\hat\nu,\mu}U^\dagger_{n,\nu})\\
    &=
    \sum_r f_r 
    \sum_{i,j,k,l}
    (U_{n,\mu})^r_{ij}
    (U_{n+\hat\mu,\nu})^r_{jk}
    (U^\dagger_{n+\hat\nu,\mu})^r_{kl}
    (U^\dagger_{n,\nu})^r_{li}
    \label{eq:character_expansion}
\end{align}
where $r$ sums over irreps (irreducible representations) of $G$ and $f_r$ is the expansion coefficient associated with $r$. The symbol $U^r_{ij}\equiv D^r_{ij}(U)$ denotes the matrix elements of $U$ under the representation $r$. See Ref. \cite{Hirasawa:2021qvh} for the parametrization of irreps of the $\text{U}(N)$ and $\text{SU}(N)$ gauge groups and how the coefficient $f_r$ is computed. The diagrammatic depiction of the $r$-character $\tr _r P_{n,\mu\nu}$ is shown in figure \ref{fig:plaquette}.

Now that the Boltzmann weight is written as a linear summation of link variables, it is now easy to integrate each link separately. For a given link variable $U_{n,\mu}$, the integral reads
\begin{equation}
    L_{n,\mu}=\int dU_{n,\mu}(U_{n,\mu})^{r_1}_{i_1i_1'}\cdots (U_{n,\mu})^{r_{d-1}}_{i_{d-1}i_{d-1}'}(U^\dagger_{n,\mu})^{s_1}_{j_1j_1'}\cdots(U^\dagger_{n,\mu})^{s_{d-1}}_{j_{d-1}j_{d-1}'}
    \label{eq:link-integral}
\end{equation}
We can consider $L_{n,\mu}$ as a tensor located on each link with the tensor indices being the representation $r_a$, $s_a$, and the matrix indices $i_a,i'_a,j_a,j'_a$.

For each plaquette, it is required that the 4 link variables must be under the same representation. Thus, we must also define a plaquette tensor
\begin{equation}
    (R_{n,\mu\nu})_{r_1r_2r_3r_4}=
    f_{r_1}
    \delta_{r_1r_2}\delta_{r_1r_3}\delta_{r_1r_4}.
    \label{eq:R-tensor}
\end{equation}

The partition function can then be written as a tensor network via
\begin{equation}
    Z = \text{tr}\prod_{n\in\Lambda_d}\left(\prod_{\mu}L_{n,\mu}\prod_{\mu<\nu}R_{n,\mu\nu}\right).
\end{equation}

\begin{figure}
    \centering
    \includegraphics[scale=0.9]{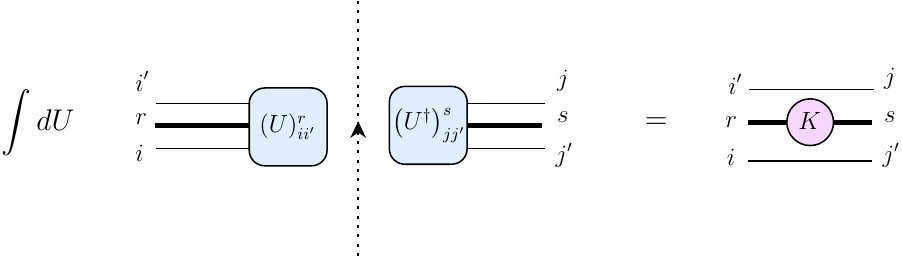}
    \caption{Diagrammatic representation of the Schur orthogonality relations \eqref{eq:got}. The dashed line with an arrow shows the orientation of the link variable in comparison with the position of the matrix indices---the indices $ii'$ in $U_{ii'}$ align in the same way as the link while the indices $jj'$ in $U^\dagger_{jj'}$ align oppositely.}{\color{white}.}\\[5mm]
    \label{fig:got}
    \includegraphics[scale=0.8]{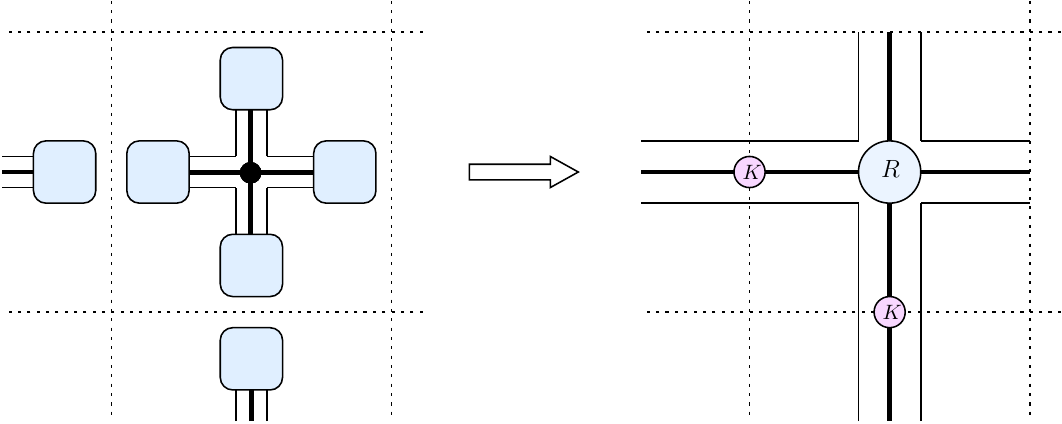}
    \caption{The tensor network representation of the two-dimensional gauge theory before (left) and after (right) the integration. The tensor $R$ and $K$, which only have representation indices (thick lines), are given in \eqref{eq:R-tensor} and \eqref{eq:K-tensor}, respectively. The matrix indices (thin lines) form a loop around each site, which can be traced out and give a factor of $\text{dim}(r)$.}
    \label{fig:2dtensor}
\end{figure}

In two dimensions, the link integral can be integrated directly using the Schur orthogonality relations
\begin{equation}
    L_{n,\mu}=\int dU_{n,\mu}(U_{n,\mu})^r_{ii'}(U^\dagger_{n,\mu})^s_{jj'}=K_{rs}\delta_{ji'}\delta_{ij'}.
    \label{eq:got}
\end{equation}
where we define the link-conditional tensor
\begin{equation}
    K_{rs}=\frac{1}{\text{dim}(r)}\delta_{rs},
    \label{eq:K-tensor}
\end{equation}
with $\text{dim}(r)$ being the dimensionality of the representation $r$. This tensor gives a restriction on possible representations of the two adjacent plaquettes.
This is depicted diagrammatically as in figure \ref{fig:got}.

The tensor network of the two-dimensional gauge theory in terms of $R$ and $K$ are shown in figure \ref{fig:2dtensor}. It is not difficult to notice that the matrix indices form a loop that can be traced out analytically, giving a factor of $\text{dim}(r)$ per site. The resulting tensor network is now completely free of the matrix indices, which is very beneficial for numerical calculations because the tensors are now significantly smaller and are free of the degeneracy in the singular value spectrum. Such a reduction was known to be particular to two dimensions. As we will show next, such a reduction is also possible in higher dimensions, albeit with a more complicated form.

\section{Index loops in higher dimensions}
\label{sec:loops}
\subsection{Generalized Clebsch-Gordan decomposition}
The link integral \eqref{eq:link-integral} in higher dimensions is more complicated since the Schur orthogonality relations does not straightforwardly apply. To perform such an integral, one needs to rewrite a tensor product of the link variable as a direct sum of irreps first. The simplest form of such a decomposition is the Clebsch-Gordan decomposition
\begin{equation}
    U_{i_1i_1'}^{r_1}U_{i_2i_2'}^{r_2} = \sum_{\hat r\in D_{r_1\otimes r_2}}\sum_{\hat\imath,\hat\imath'}C^{{\hat r}\hat\imath}_{r_1i_1;r_2i_2}C^{{\hat r}\hat\imath'}_{r_1i_1';r_2i_2'}U_{\hat\imath\hat\imath'}^{\hat r}
\end{equation}
where $D_{r_1\otimes r_2}$ is a set of irreps obtained from the decomposition of $r_1\otimes r_2$ and
\begin{equation}
    C^{{\hat r}\hat\imath}_{r_1i_1;r_2i_2}=\langle r_1i_1;r_2i_2|{\hat r}\hat\imath\rangle
\end{equation}
is the real-valued Clebsch-Gordan coefficient. See, for example, Ref. \cite{10.1063/1.3521562}, for the underlying theory and how to evaluate the Clebsch-Gordan coefficients for $\text{SU}(N)$ and $\text{SL}(N,\mathbb{C})$ groups. For a product of more than two matrices, more decomposition can be successively done to reduce all the tensor products into a direct sum of irreps; e.g.,
\begin{align}
    r_1\otimes r_2\otimes r_3 &= (r_1\otimes r_2)\otimes r_3\nonumber\\
    &= ({\hat r}_1\oplus {\hat r}_2\oplus\cdots)\otimes r_3\nonumber\\
    &= ({\hat r}_1\otimes r_3)\oplus ({\hat r}_2\otimes r_3)\oplus\cdots\nonumber\\
    &= ({\hat r}_{11}\oplus{\hat r}_{12}\cdots)\oplus ({\hat r}_{21}\oplus{\hat r}_{22}\oplus\cdots)\oplus\cdots.
\end{align}
This property allows us to define a generalized Clebsch-Gordan coefficient recursively
\begin{equation}
    C^{{\hat r}\hat\imath}_{r_1i_1;\cdots;r_ni_n}=
    \sum_{\hat\jmath}
    C^{{\hat r}\hat\imath}_{{\hat s}\hat\jmath;r_ni_n}
    C^{{\hat s}\hat\jmath}_{r_1i_1;\cdots;r_{n-1}i_{n-1}}
    \label{eq:CG1}
\end{equation}
where ${\hat r}$ is an irrep obtained from the decomposition of ${\hat s}\otimes r_n$. Let us introduce the notation
\begin{equation}
    C^{\hat r\hat\imath}_{\{r\}\{i\}}\equiv C^{{\hat r}\hat\imath}_{r_1i_1;\cdots;r_ni_n}.
    \label{eq:CG2}
\end{equation}
We note here that the CG coefficients \eqref{eq:CG1}-\eqref{eq:CG2} are uniquely defined by a sequence of decompositions that leads to $\hat r$ rather than defined by $\hat r$ alone. Therefore, in what follows, the summation over $\hat r\in D_{r_1\otimes\cdots\otimes r_n}$ would mean the summation over these sequences:
\begin{equation}
    \sum_{\hat r\in D_{\otimes r}}\equiv\sum_{\rho_1\in D_{r_1\otimes r_2}}\sum_{\rho_2\in D_{\rho_1\otimes r_3}}\cdots \sum_{\rho_{n-1}\in D_{\rho_{n-2}\otimes r_n}},
\end{equation}
where $\rho_{n-1}\equiv\hat r$ is the final representation of the sequence.

\subsubsection*{Some comments on the generalized Clebsch-Gordan coefficients}

The coefficient $C^{\hat r\hat\imath}_{\{r\}\{i\}}$ can be considered as a one-to-one linear map of a vector space from the `uncoupled' basis $|\{r\}\{i\}\rangle$ to the `coupled' basis $|\hat r\hat\imath\rangle$. Certainly, these bases span the same vector space and satisfy their own orthogonality relation
\begin{align}
    \langle \{r\}\{i\}|\{r\}\{i'\}\rangle&=\prod_{a=1}^n\delta_{i_ai'_a},\\
    \langle \hat r\hat\imath|\hat r'\hat\imath'\rangle&=\delta_{\hat r\hat r'}\delta_{\hat\imath\hat\imath'}.
\end{align}
Since $C^{\hat r\hat\imath}_{\{r\}\{i\}}\equiv\langle \{r\}\{i\}|\hat r\hat\imath\rangle$, this implies that the CG coefficients are orthogonal
\begin{align}
    \sum_{\hat r,\hat\imath}C^{\hat r\hat\imath}_{\{r\}\{i\}}C^{\hat r\hat\imath}_{\{r\}\{i'\}}
    &=\prod_{a=1}^n\delta_{i_ai'_a},\\
    \sum_{\{i\}}C^{\hat r\hat\imath}_{\{r\}\{i\}}C^{\hat r'\hat\imath'}_{\{r\}\{i\}}&=
    \delta_{\hat r\hat r'}\delta_{\hat\imath\hat\imath'}.
\end{align}
This orthogonality also holds even if the two coupled representations are equivalent\footnote{It is common that an irrep $\hat r_a$ obtained from the CG decomposition of $r_1\otimes\cdots\otimes r_{d-1}$ occur multiple times. Such representations are thus usually given an additional multiplicity index. In this paper, unless explicitly stated, we will treat all of these irreps to be unique, even if some of them are equivalent.}, $\hat r\simeq\hat r'$, because they span subspaces that are orthogonal to each other.

The CG coefficients can always be chosen to be real, but even so, they are still not completely unique due to the $\text{SO}(\text{dim}(\hat r))$ symmetry of the basis transformation
\begin{equation}
    C^{\hat r\hat\imath}_{\{r\}\{i\}}\rightarrow \sum_{\hat\imath'}C^{\hat r\hat\imath'}_{\{r\}\{i\}}\Omega^{\hat r}_{\hat\imath'\hat \imath}.
\end{equation}
In this paper, we will fix the basis transformation matrix $\Omega^{\hat r}$ to be the one such that the link variable $U\in G$ has the same matrix elements in every equivalent representation:
\begin{equation}
    U^{\hat r}-U^{\hat r'}=0\quad\text{for all}\quad\hat r\simeq\hat r'.
\end{equation}

In what follows, we define
\begin{equation}
    \delta(\hat r\simeq\hat s) \equiv \left\{
    \begin{array}{ll}
    1 &;\quad \hat r\simeq\hat s,\\
    0 &;\quad\text{otherwise},
    \end{array}
    \right.
\end{equation}
which is a weaker version of $\delta_{\hat r\hat s}$ that is nonzero only if $r$ and $s$ are of the same vector space.
\subsection{The armillary sphere}

Let us once again consider the link integral \eqref{eq:link-integral}
\begin{equation}
    L_{n,\mu}=\int dU_{n,\mu}(U_{n,\mu})^{r_1}_{i_1i_1'}\cdots (U_{n,\mu})^{r_{d-1}}_{i_{d-1}i_{d-1}'}(U^\dagger_{n,\mu})^{s_1}_{j_1j_1'}\cdots(U^\dagger_{n,\mu})^{s_{d-1}}_{j_{d-1}j_{d-1}'}
    \label{eq:link-integral2}
\end{equation}
We now assume the following complete decomposition
\begin{align}
    r_1\otimes\cdots\otimes r_{d-1}&={\hat r}_1\oplus\cdots\oplus{\hat r}_p,\label{eq:r-decomposition}\\
    s_1\otimes\cdots\otimes s_{d-1}&={\hat s}_1\oplus\cdots\oplus{\hat s}_q.\label{eq:s-decomposition}
\end{align}
with ${\hat r}_a$ and ${\hat s}_b$ being irreps of $G$. We define the set $D_{\otimes r}=\{{\hat r}_1,{\hat r}_2,\cdots,{\hat r}_p\}$ and $D_{\otimes s}=\{{\hat s}_1,{\hat s}_2,\cdots,{\hat s}_q\}$ to denote the sets of the decomposed irreps.
In terms of the matrix elements, we have
\begin{align}
    (U_{n,\mu})^{r_1}_{i_1i_1'}\cdots (U_{n,\mu})^{r_{d-1}}_{i_{d-1}i_{d-1}'} &= \sum_{{\hat r}\in D_{\otimes r}}
    \sum_{\hat\imath,\hat\imath'}
    C^{\hat r\hat\imath}_{\{r\}\{i\}}
    C^{\hat r\hat\imath'}_{\{r\}\{i'\}}
    (U_{n,\mu})^{\hat r}_{\hat\imath\hat\imath'},\\
    (U^\dagger_{n,\mu})^{s_1}_{j_1j_1'}\cdots(U^\dagger_{n,\mu})^{s_{d-1}}_{j_{d-1}j_{d-1}'}&= \sum_{{\hat s}\in D_{\otimes s}}
    \sum_{\hat\jmath,\hat\jmath'}
    C^{\hat s\hat\jmath}_{\{s\}\{j\}}
    C^{\hat s\hat\jmath'}_{\{s\}\{j'\}}
    (U^\dagger_{n,\mu})^{\hat r}_{\hat\jmath\hat\jmath'}.
\end{align}
With this decomposition, the link integral becomes
\begin{align}
    L_{n,\mu}&=
    \sum_{{\hat r}\in D_{\otimes r}}\sum_{{\hat s}\in D_{\otimes s}}
    \sum_{\hat\imath,\hat\imath',\hat\jmath,\hat\jmath'}
    C^{\hat r\hat\imath}_{\{r\}\{i\}}
    C^{\hat r\hat\imath'}_{\{r\}\{i'\}}
    C^{\hat s\hat\jmath}_{\{s\}\{j\}}
    C^{\hat s\hat\jmath'}_{\{s\}\{j'\}}
    \int dU_{n,\mu}
    (U_{n,\mu})^{\hat r}_{\hat\imath\hat\imath'}
    (U_{n,\mu}^\dagger)^{\hat s}_{\hat\jmath\hat\jmath'}
    \nonumber
    \\
    &=
    \sum_{{\hat r}\in D_{\otimes r}}\sum_{{\hat s}\in D_{\otimes s}}
    \sum_{\hat\imath,\hat\jmath}
    \frac{1}{\text{dim}(\hat r)}
    C^{\hat r\hat\imath}_{\{r\}\{i\}}
    C^{\hat r\hat\jmath}_{\{r\}\{i'\}}
    C^{\hat s\hat\jmath}_{\{s\}\{j\}}
    C^{\hat s\hat\imath}_{\{s\}\{j'\}}
    \delta(\hat r\simeq \hat s).
\end{align}
We now define the vertex tensor and the link-conditional tensor
\begin{align}
    V^\alpha_{\{i\}\{j'\}}&=
    \delta(\hat r\simeq \hat s)
    \sum_{\hat\imath}
    C^{\hat r\hat\imath}_{\{r\}\{i\}}
    C^{\hat s\hat\imath}_{\{s\}\{j'\}},
    \label{eq:vertex-tensor}\\
    K^{\alpha\beta}_{\{r\}\{s\}} &= 
    \frac{1}{\text{dim}(\hat r)}
    \delta_{\hat r\hat r'}
    \delta_{\hat s\hat s'}
    \delta(\hat r\in D_{\otimes r})
    \delta(\hat s\in D_{\otimes s});
    \label{eq:ndK}\\
    \alpha&=(\{r\},\{s\};\hat r,\hat s),\\
    \beta&=(\{r\},\{s\};\hat r',\hat s'),
\end{align}
where
\begin{equation}
    \delta(\hat r\in D_{\otimes r}) = \left\{
    \begin{array}{ll}
    1&;\quad\hat r\in D_{\otimes r},\\
    0&;\quad\text{otherwise}.
    \end{array}
    \right.
\end{equation}
The link integral can then be written compactly as
\begin{equation}
    L_{n,\mu}=
    \sum_{\alpha,\beta}
    K^{\alpha\beta}_{\{r\}\{s\}}
    V^{\alpha}_{\{i\}\{j'\}}
    V^{\beta}_{\{i'\}\{j\}}.
    \label{eq:L-V}
\end{equation}
Note that the summation over multi-representation indices $\alpha,\beta$ here actually means the summation over the last two components only; e.g., $\hat r$ and $\hat s$ for summing over $\alpha$. Let us also stress again that $\delta_{\hat r\hat r'}$ requires $\hat r$ and $\hat r'$ to be of the same vector space while $\delta(\hat r\simeq\hat r')$ also allows for representations from two isomorphic but orthogonal subspaces.

\begin{figure}
    \centering
    \includegraphics[scale=0.8]{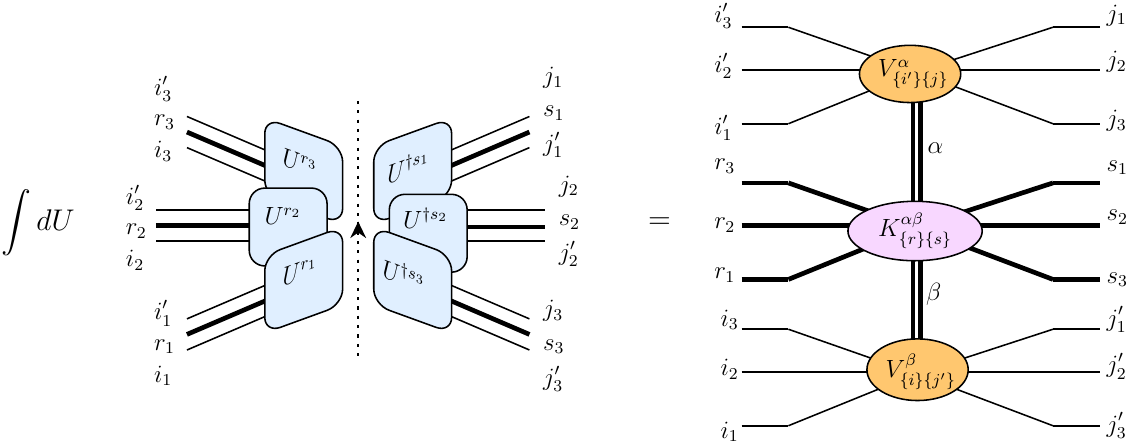}
    \caption{The diagrammatic representation of the link integral $L_{n,\mu}$ (\eqref{eq:link-integral2} and \eqref{eq:L-V}). Thin lines are matrix indices while the thick lines are representation indices. The vertical double lines are the multi-representation indices. The example above is with $d=4$.
    Similarly to the two-dimensional case, the indices $ii'$ in $U_{ii'}$ align in the same way as the link (the arrow) while the indices $jj'$ in $U^\dagger_{jj'}$ align oppositely.
    }{\color{white}.}\\[5mm]
    \label{fig:link-integral}
    \includegraphics[scale=0.7]{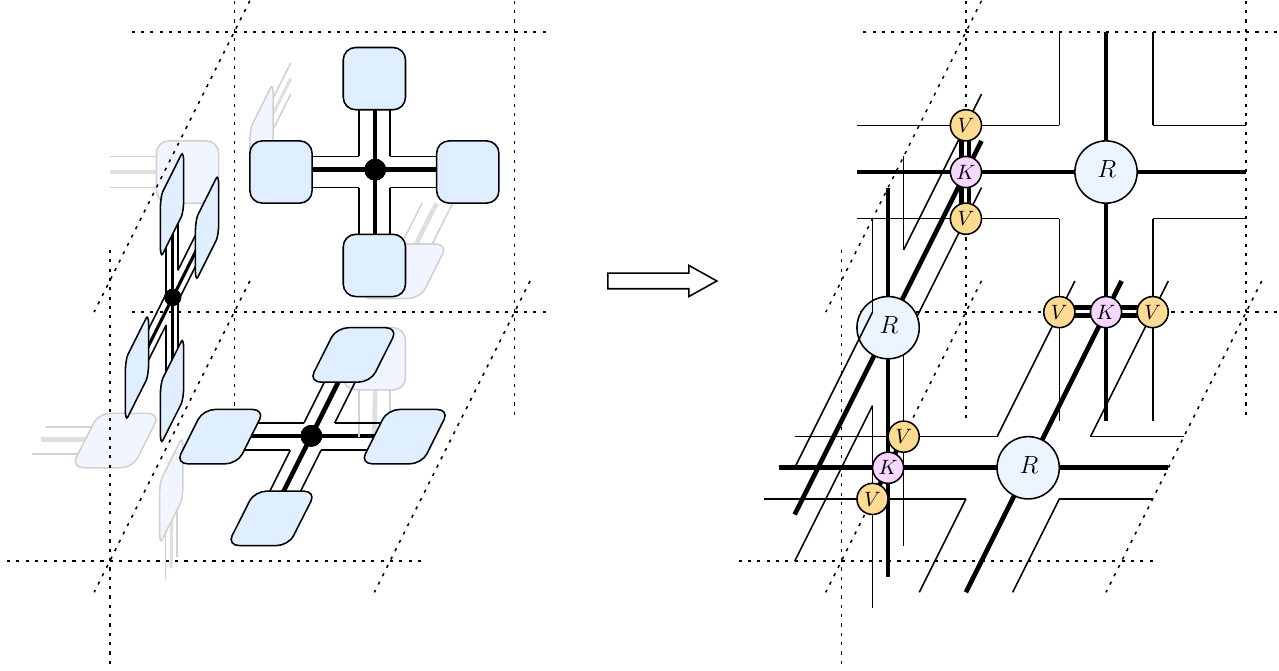}
    \caption{The tensor network representation of the three-dimensional gauge theory before (left) and after (right) the integration. The tensor $R$, and $K$, which only have representation indices, are given in \eqref{eq:R-tensor} and \eqref{eq:ndK}, respectively. The only tensors with matrix indices are the vertex tensors $V$ \eqref{eq:vertex-tensor}.}
    \label{fig:3dtensor}
\end{figure}

Equation \eqref{eq:L-V} can be written as a tensor diagram in figure \ref{fig:link-integral}.
The three-dimensional example of the tensor network is shown in figure \ref{fig:3dtensor}.

Notice that the vertex tensors surrounding a particular site actually form a special network where every matrix indices are completely contracted, leaving only the representation indices. We refer to this kind of structure as the \emph{armillary sphere}. This is the higher-dimensional generalization of the index loop in two dimensions. Once the matrix indices are completely contracted, we obtain the armillary tensor at site $n$
\begin{equation}
    A^{\{\alpha\}}_n \equiv \tr \prod_\text{vertices} V^\alpha_{\{i\}\{j\}}.
    \label{eq:armillary}
\end{equation}
The armillary tensor is completely group-theoretical in the sense that it only depends on the space-time geometry and not on physical parameters.

\subsubsection*{Three-dimensional example}
\begin{figure}
    \centering
    \includegraphics[scale=0.8]{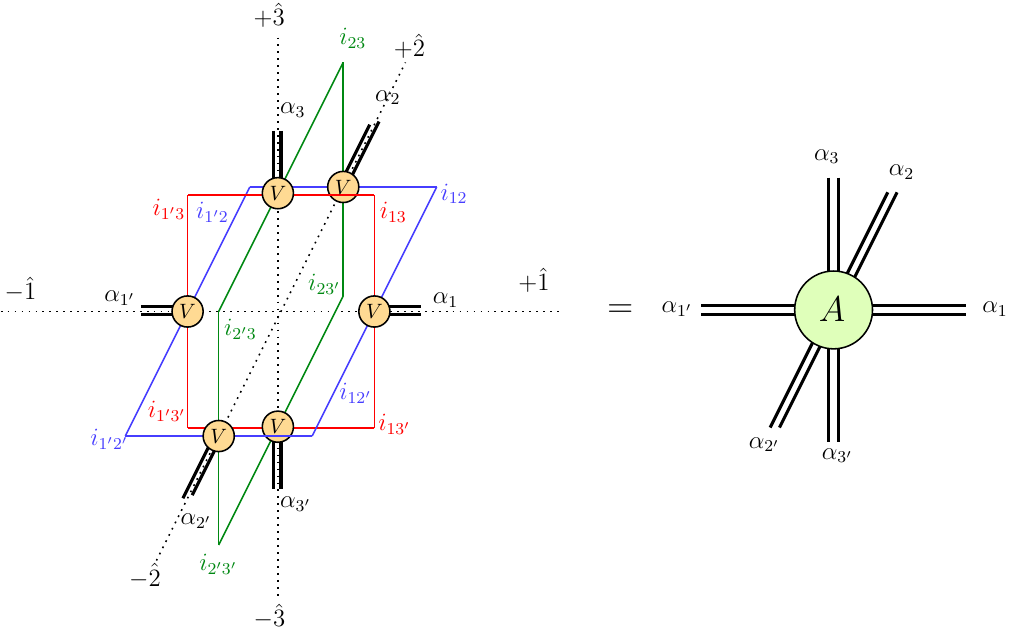}
    \caption{(Color online) Three-dimensional armillary sphere \eqref{eq:3darmillary}. Matrix legs in different planes are colored differently.}
    \label{fig:armillary}
\end{figure}

\begin{figure}
    \centering
    \includegraphics[scale=0.7]{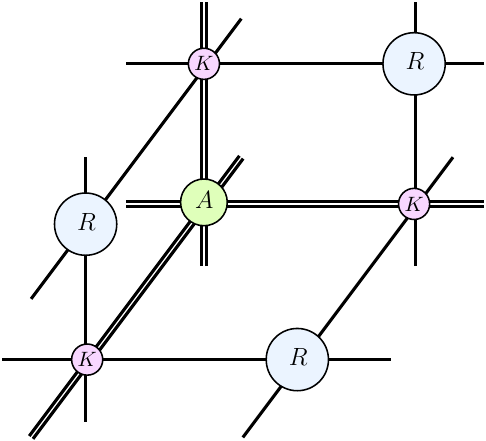}
    \caption{The reduced tensor network in three dimensions. Every leg shown is the representation leg.}
    \label{fig:3dtensor_reduced}
\end{figure}

Let $r_{\mu\nu}$, $r_{\mu'\nu}$, $r_{\mu\nu'}$, and $r_{\mu'\nu'}$ with $\mu,\nu=1,2,3$ and $\mu<\nu$ be the representation associated with the plaquette $P_{n,\mu\nu}$, $P_{n-\hat\mu,\mu\nu}$, $P_{n-\hat\nu,\mu\nu}$, and $P_{n-\hat\mu-\hat\nu,\mu\nu}$, respectively. The index $i_{\mu\nu}$ (and $i_{\mu'\nu}$, etc.) refers to the matrix index under the representation $r_{\mu\nu}$ (and $r_{\mu'\nu}$, etc.). See figure \ref{fig:armillary} for the location of the indices and the vertices. The six vertex tensors are
\begin{itemize}
    \item $V^{\alpha_1}_{i_{12}i_{13}i_{12'}i_{13'}}$;
    $\alpha_{1}=(r_{12},r_{13},r_{12'},r_{13'};\hat r_{1},\hat s_{1})$,
    located on the link $+\hat 1$
    \item $V^{\alpha_{1'}}_{i_{1'2}i_{1'3}i_{1'2'}i_{1'3'}}$;
    $\alpha_{1'}=(r_{1'2},r_{1'3},r_{1'2'},r_{1'3'};\hat r_{1'},\hat s_{1'})$,
    located on the link $-\hat 1$
    \item $V^{\alpha_2}_{i_{12}i_{23}i_{12'}i_{23'}}$;
    $\alpha_{2}=(r_{12},r_{23},r_{12'},r_{23'};\hat r_{2},\hat s_{2})$,
    located on the link $+\hat 2$
    \item $V^{\alpha_{2'}}_{i_{1'2}i_{2'3}i_{1'2'}i_{2'3'}}$;
    $\alpha_{2'}=(r_{1'2},r_{2'3},r_{1'2'},r_{2'3'};\hat r_{2'},\hat s_{2'})$,
    located on the link $-\hat 2$
    \item $V^{\alpha_3}_{i_{13}i_{23}i_{13'}i_{23'}}$;
    $\alpha_{3}=(r_{13},r_{23},r_{13'},r_{23'};\hat r_{3},\hat s_{3})$,
    located on the link $+\hat 3$
    \item $V^{\alpha_{3'}}_{i_{1'3}i_{2'3}i_{1'3'}i_{2'3'}}$;
    $\alpha_{3'}=(r_{1'3},r_{2'3},r_{1'3'},r_{2'3'};\hat r_{3'},\hat s_{3'})$,
    located on the link $-\hat 3$
\end{itemize}
We can then define the three-dimensional armillary tensor as
\begin{align}
    A^{\alpha_{1}\alpha_{1'}\alpha_{2}\alpha_{2'}\alpha_{3}\alpha_{3'}}_n&=
    \sum_{\{i\}}
    V^{\alpha_1}_{i_{12}i_{13}i_{12'}i_{13'}}
    V^{\alpha_{1'}}_{i_{1'2}i_{1'3}i_{1'2'}i_{1'3'}}
    V^{\alpha_2}_{i_{12}i_{23}i_{12'}i_{23'}}
    \nonumber\\
    &\;\;\;\;\;\;\;\;\;\;\;\;\;\;\;\;\;\;\;\;\;\;\;\;\;\;\;\;\;
    \times
    V^{\alpha_{2'}}_{i_{1'2}i_{2'3}i_{1'2'}i_{2'3'}}
    V^{\alpha_3}_{i_{13}i_{23}i_{13'}i_{23'}}
    V^{\alpha_{3'}}_{i_{1'3}i_{2'3}i_{1'3'}i_{2'3'}},
    \label{eq:3darmillary}
\end{align}
as shown in figure \ref{fig:armillary}. Figure \ref{fig:3dtensor_reduced} shows the resulting tensor network on one lattice site where every leg corresponds to the (multi-)representation index.

\subsection{More general cases}

\begin{figure}
    \centering
    \includegraphics{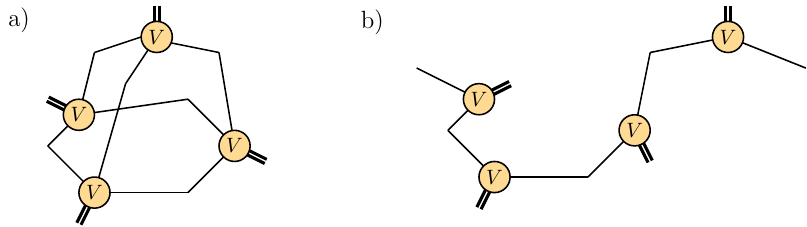}
    \caption{Two possible scenarios: a) The matrix indices form a complete loop within an armillary sphere. In this case, the matrix indices can be completely eliminated via the complete contraction; b) The matrix indices form an extended structure on the lattice. In this case, we cannot completely eliminate the matrix indices.}
    \label{fig:vertex_structure}
\end{figure}

The discussion can be directly extended to systems where the action is a function of more complicated Wilson loops $W$. Consider an action of the form
\begin{align}
    S_W &= \lambda\tr W;\\
    W&=U_1U_2\cdots U_\ell\in G,
\end{align}
where the link variables in $W$ are connected into a loop. The associated Boltzmann weight can be expanded in terms of characters as
\begin{equation}
    e^{\lambda\tr W} = \sum_r g_r (U_1)^{r}_{i_1i_2}(U_2)^{r}_{i_2i_3}\cdots (U_\ell)^{r}_{i_\ell i_1}.
\end{equation}
Each of the link variables here is then integrated together with those from the plaquette action. The resulting vertex tensors $V$ and the link-conditional tensor $K$ will have additional legs, but the separation of the indices into two layers is the same. The important question is if the new legs also belong to the armillary sphere or if they form an open structure on the lattice (see figure \ref{fig:vertex_structure}).

To answer the question, note that the matrix index originates from the matrix product between two link variables. In order for the Wilson loop $W$ to be gauge invariant, the two link variables must connect at some site $n$. This allows us to associate any matrix leg with some lattice site $n$, which in turn guarantees that it always belongs to an armillary sphere located at $n$. Figure \ref{fig:cloverleaf} shows an example in the case of the four-dimensional topological charge (the clover Wilson loop) where every matrix index clearly belongs to some armillary sphere.

\begin{figure}
    \centering
    \includegraphics[scale=0.6]{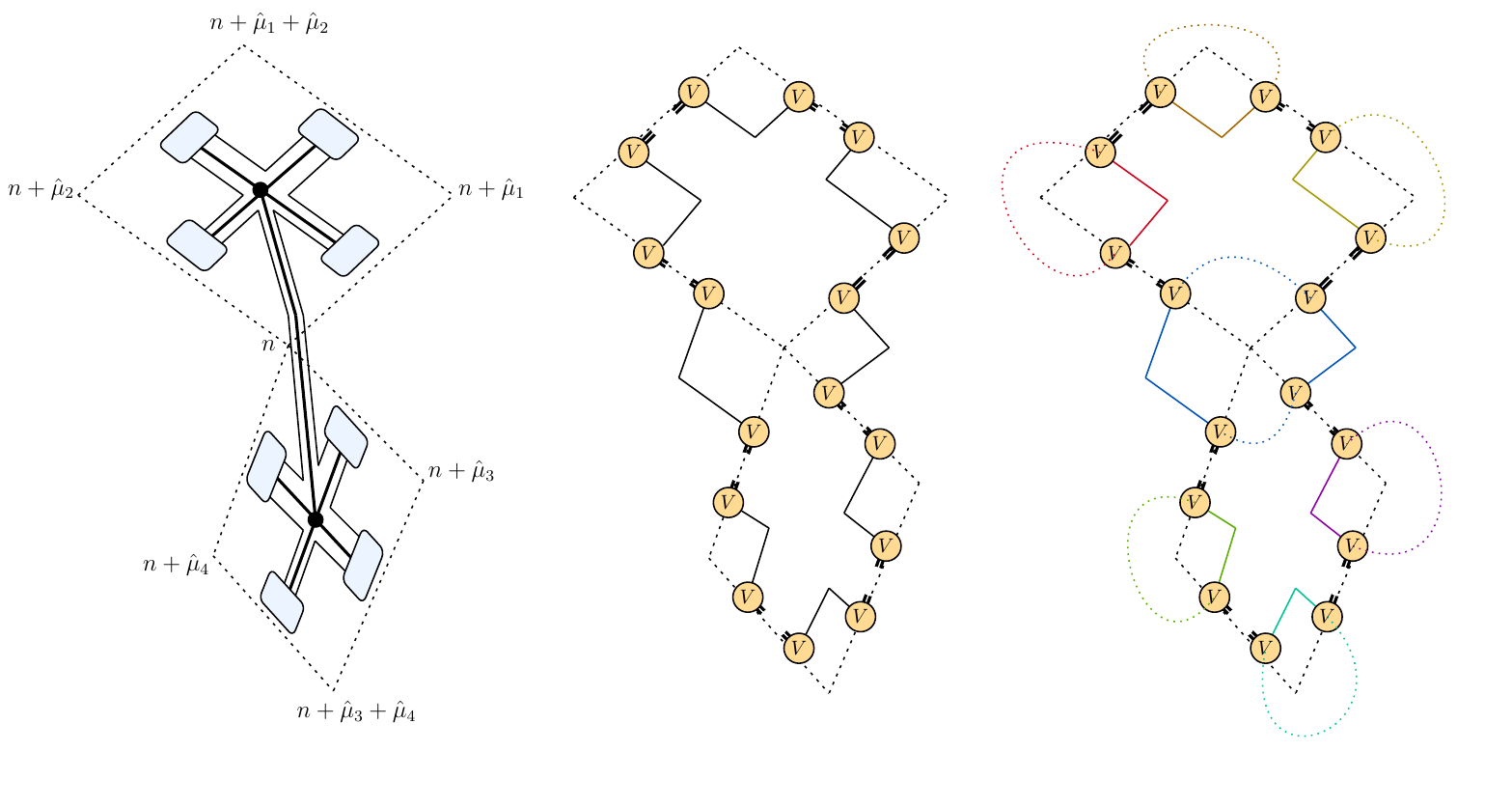}
    \caption{Left: The diagrammatic depiction of the clover Wilson loop. Middle: Vertex tensors obtained from integrating the link variables with the matrix indices shown as the solid lines. (Link-conditional tensors are omitted.) Right: Each of the matrix indices is part of an armillary sphere centered at some lattice site. The additional dotted curves show the outline of various corresponding armillary spheres.}
    \label{fig:cloverleaf}
\end{figure}

\section{Summary and discussion}
\label{sec:summary}

In this work, we proposed a new tensor network formulation for non-Abelian gauge theories in arbitrary dimensions where the matrix degrees of freedom are completely eliminated. The idea is to rewrite the integral of the link variables as two vertex tensors. Such vertex tensors can be shown to always form a closed structure around each site, which we refer to as the armillary sphere, where the matrix indices are contracted completely, leaving only the representation indices in the tensor network. Such a technique is then shown to also work for actions that are a trace of more general Wilson loops.

The upshot of this technique is that the constituent tensors on one site now become smaller while including the same amount of representations. It is also expected that the singular value spectrum will no longer have the degeneracy associated with the non-Abelian internal degrees of freedom. However, the evaluation of the armillary tensor itself can be numerically challenging since it involves a large number of matrix multiplications and the evaluation of various Clebsch-Gordan coefficients. We plan to address this issue through further investigation. The numerical computation for the 2+1D SU(2) and SU(3) gauge theories is undergoing, whose results will be reported elsewhere. Another interesting direction is to generalize the idea for theories with matter fields. We have high hopes that this work will contribute significantly to advancing the computation of gauge theory in higher dimensions.

\section*{Acknowledgments}
This work is supported by a Grant-in-Aid for Transformative Research Areas
``The Natural Laws of Extreme Universe—A New Paradigm for Spacetime and Matter from
Quantum Information” (KAKENHI Grant No. JP21H05191) from JSPS of Japan.

\bibliographystyle{JHEP}
\bibliography{ref}

\end{document}